# Effect of Nitrogen on the Growth of Boron Doped Single Crystal Diamond

Sunil K Karna[*1], Yogesh K Vohra[1,] Patrick Kung[2] and Samuel T. Weir[3]

[1]Department of Physics, University of Alabama at Birmingham (UAB), Birmingham, AL , USA

[2]Department of Electrical and Computer Engineering, UA, Tuscaloosa, AL 35487, USA

[3]L-041, Lawrence Livermore National Lab., Livermore, CA 94550, USA

[*1]skkarna@uab.edu;

### Abstract

Boron-doped single crystal diamond films were grown homoepitaxially on synthetic (100) Type Ib diamond substrates using microwave plasma assisted chemical vapor deposition. A modification in surface morphology of the film with increasing boron concentration in the plasma has been observed using atomic force microscopy. Use of nitrogen during boron doping has been found to improve the surface morphology and the growth rate of films but it lowers the electrical conductivity of the film. The Raman spectra indicated a zone center optical phonon mode along with a few additional bands at the lower wavenumber regions. The change in the peak profile of the zone center optical phonon mode and its downshift were observed with the increasing boron content in the film. However, shrinkage and upshift of Raman line was observed in the film that was grown in presence of nitrogen along with diborane in process gas.

### Introduction

Doped diamond films are of interest because of their potential use in the semiconductor industry as well as other applications involving their unique thermal, mechanical and chemical properties. Doping of diamond with boron during its growth process is a widely investigated field of research [1-4]. Boron provides diamond a wide range of dopant concentrations ($10^{16}$ up to $10^{21}$ atom cm-3) from a wide band gap semiconductor to a metal and to a superconductor [5-7]. However, the large scale production of doped diamond film can be possible only if high growth rate and high crystalline quality can be achieved simultaneously. It has been reported that the use of nitrogen during deposition improves the growth rate of diamond films [8-10]. In this study we have investigated the effect of nitrogen on growth morphology and growth rate of diamond films during boron doping and also studied their electrical conductivity as a function of temperature.

### Experimental

Synthetic (100) oriented Type Ib diamond substrates (size 3.5 × 3.5 × 1.5 mm$^3$) were chosen as seed crystals for this study. The acetone cleaned seed crystal was inserted into 2.45 GHz microwave reactor chemical vapor deposition (CVD) chamber. The microwave power was adjusted to 1.4-1.5 kW to set a deposition temperature of 1100 ± 20º C at a chamber pressure of 100 Torr. The 10% $B_2H_6$ diluted in $H_2$ with 6% of $CH_4$ / $H_2$ mixture was used for the deposition with a total of 400 standard cubic centimeters per minute (sccm) of gas. The details of deposition parameters and growth rate of samples have been summarized in Table 1. Each sample was ultrasonicated in acetone after deposition to remove any residual boron-carbon soot. The quality and surface morphology of as deposited films were determined by Raman spectroscopy, X-ray rocking curve experiment, optical microscopy (OM) and atomic force microscopy (AFM). Raman spectra were recorded using a 514 nm laser excitation wavelength at room temperature. In the X-ray rocking curve experiment, omega scans were obtained by rotating sample with 0.02º angular step with a detector fixed at 2θ position corresponding to (400) Bragg diffraction peak. Type of doping, doping level and electrical conductivity was determined by room temperature Hall measurement and four point probe measurements respectively. In a four point probe measurement, the samples were first heated to 600K and the electrical resistance was measured during the cool down process. This procedure was followed to remove any adsorbates from the diamond films that could have arisen from the CVD shut down procedure in hydrogen atmosphere. The I-V characteristics were maintained in the linear regime during experiment by limiting the current supply within 1mA at zero – magnetic field. The value of sheet resistance was measured in magnetic field of 0.55 Tesla. The measurements were taken for both positive and negative current with both polarities of magnetic field.

TABLE 1: GROWTH CONDITIONS, GROWTH RATE [r] AND BULK CARRIER CONCENTRATIONS [η] OF GROWN DIAMOND FILMS.

| Sample | B2H6 (ppm) | [B/C]gas (ppm) | N2 (ppm) | [r] (μm/hr) | [η] (cm$^{-3}$) |
|---|---|---|---|---|---|
| HD2 | 0 | 0 | 0 | 10 | 5.5×10$^{17}$ |
| (BD5) | 150 | 5000 | 0 | 6 | 1.0×10$^{19}$ |
| (BD6) | 250 | 8000 | 0 | 5 | 6.6×10$^{19}$ |
| (BD7) | 500 | 16000 | 0 | 4 | 2.3×10$^{20}$ |
| (BD8) | 500 | 16000 | 1000 | 6 | 3.9×10$^{19}$ |
| (BD9) | 500 | 16000 | 2000 | 12 | - |

## Results and discussion

In order to study the effect of nitrogen on growth morphology of boron doped diamond, a few samples were grown at varying diborane and nitrogen contents in the feed gas for 5 hours deposition. After doping the yellow color of the seed changed into pale blue, then to dark blue and finally opaque to visible light due to the increasing concentration of boron in the films. Optical transmission spectra of the samples in the visible range were taken in order to measure the effect of doping on the color of the films as shown in Fig. 1. The spectra indicate small downshift of peak and decrease in intensity as the doping level of the films increases. The peak of low doped film (sample BD5) appeared at 454 nm in the optical transmission spectra (Fig. 1). The surface morphology of films was analysed using OM and AFM images taken over scanned areas of 10 × 10 μm$^2$ as shown in Fig. 2 (a, b and c). The untreated substrate samples (seeds) are macroscopically flat. The OM image of undoped diamond film (HD2) shows some hillocks on its surface as shown in Fig. 2(a). The growth rate and surface morphology are observed to be dependent on B/C ratio in the gas phase. Increasing B/C ratio up to a certain limit in feed gas decreases the growth rate as well as the surface roughness. The surface smoothness of the film was improved with an additional supply of 1000 ppm of nitrogen in feed gas as shown by a comparison of sample BD7 (16000 ppm of B2H6 and no N2) and BD8 (addition of 1000 ppm of N2) in Fig. 2(b) and Fig. 2(c). However, hillocks were observed on the surface of film with the addition of 2000 ppm of nitrogen (BD9), but it improved the growth rate of diamond films by a factor two compared to that using 1000 ppm of nitrogen as shown in Table 1 of samples BD8 and BD9. The root mean square (rms) roughness of the samples measured from contact mode AFM images of the surface are listed in parenthesis for each sample Seed (1.6 nm), HD2 (2nm), BD5 (4nm), BD6 (1.6nm), BD7 (1nm), BD8 (0.8nm) and BD9 (0.9 nm) and was found to decrease with increasing boron concentration. Introduction of 1000 ppm of nitrogen during boron doping improved the surface morphology and increased the growth rate of film (sample BD8). However, when higher amount of nitrogen (2000 ppm) along with the same B/C ratio (16000 ppm) was used, the film growth rate was shown to increase by twice than that using 1000 ppm of nitrogen, but the surface quality degraded. When very high B/C ratio (25000 ppm) or very high methane 10% of total feed gas was introduced in the deposition chamber, amorphous carbon and polycrystalline diamond film grew on the seed crystal. Hence, it can be argued that an optimal amount of diborane and nitrogen is needed to improve the structural quality and the growth rate of boron doped single crystal diamond.

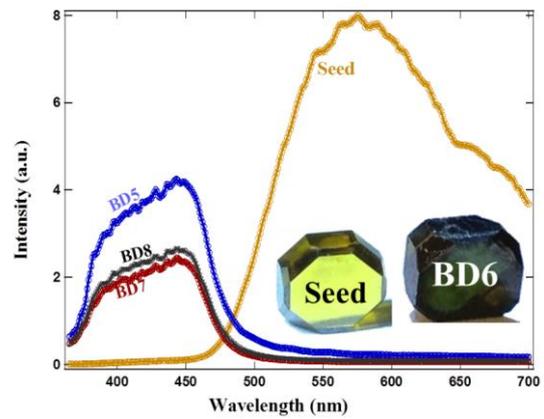

Fig. 1. Optical transmission spectra of seed crystal (Seed) and doped diamond films after applying correction for the absorption in the diamond seeed crystal. Photographs of seed crystal (Seed) and doped diamond (BD6) are embedded inside the graph. Yellow color is avidence of substitutional nitrogen atom in seed crystal and bluish color in BD6 represents presence of boron in diamond lattice.

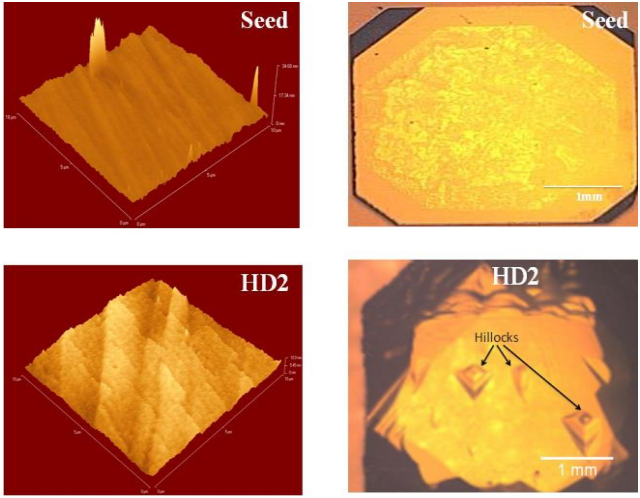

Fig. 2(a). Atomic Force Microscopy (AFM-left) and Optical Microscopy (OM-right) images of untreated substrate (Seed) and as deposited undoped diamond film (HD2).

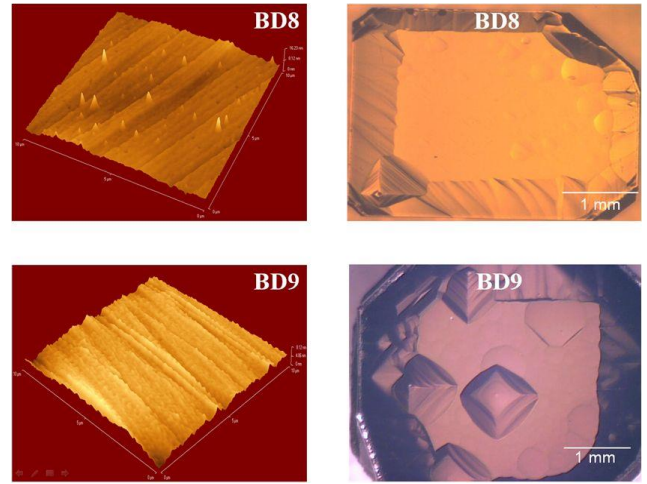

Fig. 2(c). Atomic Force Microscopy (AFM-left) and Optical Microscopy (OM-right) images of as deposited boron doped diamond films. Films BD8 and BD9 were grown in presence of 1000 and 2000 ppm of nitrogen respectively in the feed gas.

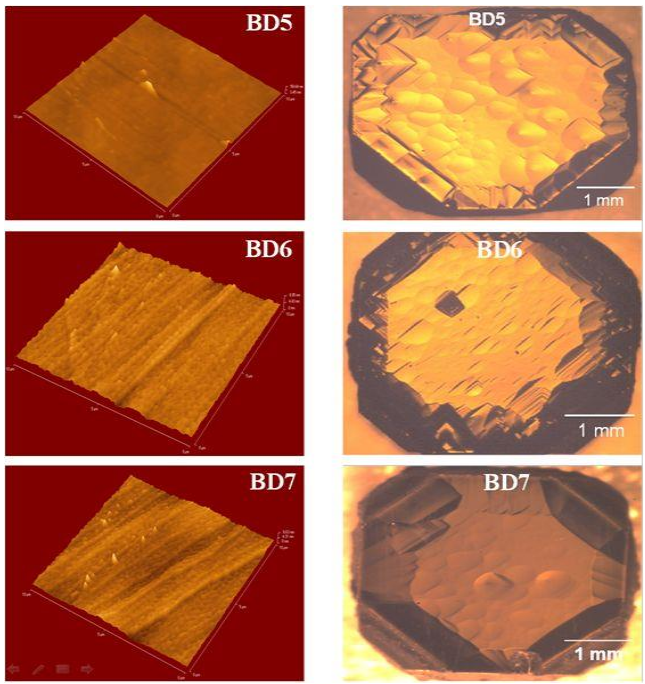

Fig. 2(b). Atomic Force Microscopy (AFM-left) and Optical Microscopy (OM-right) images of as deposited boron doped diamond films (BD5, BD6 and BD7).

The quality of the epitaxial diamond film determined using X-ray rocking curve and Raman spectroscopy is shown in Fig. 3. A single intense peak of (400) was observed in the rocking curve experiment from 30–62° omega scans indicates high quality film as shown in the inset of Fig. 3. The measured full width at half maximum (FWHM) of the Bragg peak varies from 0.07° to 0.11° as the boron content in the films is increased indicating the decrease in structural quality.

An intense zone-center optical phonon mode of diamond is visible at 1333 cm$^{-1}$ along with additional bands at 580, 900, 1042, 1233 cm$^{-1}$ as shown in Fig. 3. Significant modification and asymmetry in zone center optical phonon line was observed in the presence of such additional bands. In low boron doped films, these bands were absent and no asymmetry in Raman line was observed. Those additional bands were previously reported on both the boron doped polycrystalline and single crystal diamond [6, 11-13]. The asymmetry in optical phonon Raman line increased with increasing doping level. The downshift of optical phonon line and broadening of FWHM with doping level were also observed in the spectrum. The Raman line of boron doped samples is observed to be sharpened and shifted towards higher wavenumber region when the sample was grown in presence of nitrogen. The downshift of optical phonon line with increasing boron content has already been described by Gheeraert et al [14]. The bands around 1233 and 580 cm$^{-1}$ are assigned due to large cluster of boron atoms present in diamond lattice [7, 11, 15]. The asymmetric broadening and downshift of Raman peak of boron – doped diamond films could be explained by Fano-effect [16-20].

The variation in conductivity, (σ) of the films with temperature (T) was determined from four point probe measurement as shown in Fig. 4. Activation energies of the samples have been obtained by the best fit of the Arrhenius plot of the conductivity data in two different temperature regions. The activation energies of samples BD6, BD7 and BD8 are 0.18, 0.10 and 0.16 eV above transition point and 0.02, 0.02, 0.01 eV below transition point respectively. The difference in activation energy from high to low temperature indicates that two different conduction mechanisms are responsible for carrier transport in the film [21, 22]. At higher temperatures, carriers are transported via band conduction and at low temperatures, carrier hops in the localized states via hopping conduction. At high temperature the activation energy of carriers was found to decrease with increasing doping concentration. Increasing doping concentration increases the acceptor band width which ultimately reduces the activation energy of acceptors [20]. P-type doping was verified by Hall measurement. The boron concentrations in samples BD6, BD7 and BD8 calculated from room temperature Hall measurement were $6.6 \times 10^{19}$, $2.3 \times 10^{20}$ and $3.9 \times 10^{19}$ cm$^{-3}$ respectively and that from activation energy above transition point using Pearson and Barden formula were $2.0 \times 10^{19}$, $0.6 \times 10^{20}$ and $2.7 \times 10^{19}$ cm$^{-3}$ respectively [23].
shown in Fig. 1.

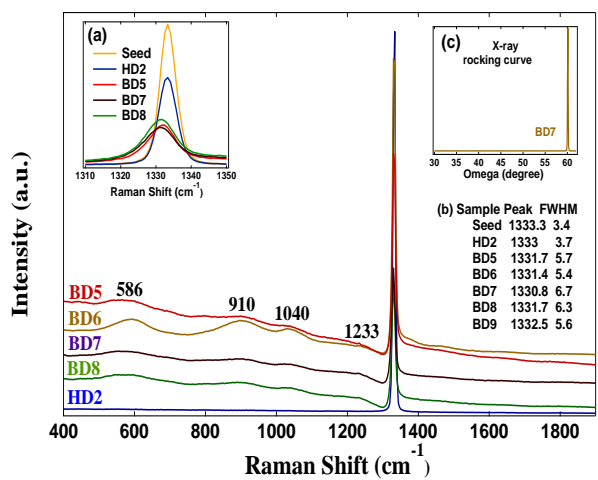

Fig. 3. Raman spectra of Seed, undoped (HD2) and doped (BD5, BD6, BD7 and BD8) diamond films. The spectra are offset for clarity. The insert (a) and (b) provide peak location of Raman spectrum and Full Width at Half Maximum (FWHM) for the zone center Raman mode. The insert (c) shows the X-ray rocking curve for sample BD7.

## Conclusions

Boron–doped single crystal diamonds have been synthesized using MPCVD method and effect of nitrogen during the growth process of boron doped diamond has also been studied. Supply of 1000 ppm of nitrogen in feed gas improves the surface morphology and increases the growth rate of diamond deposition.

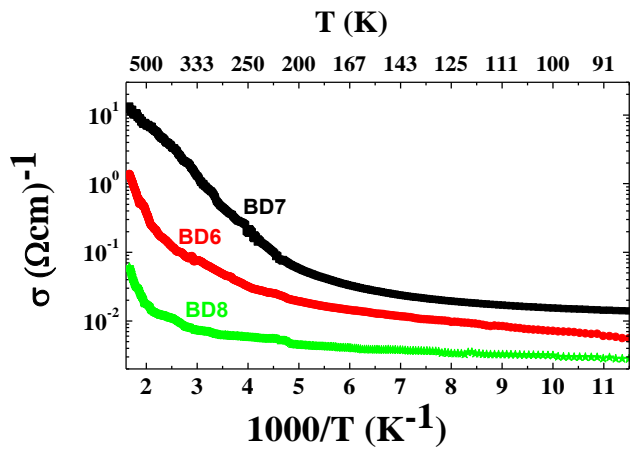

Fig. 4. Four point probe measurement of electrical conductivity of boron doped diamond films BD6, BD7 and BD8 as a function of temperature. Arrow heads are indicating the transition point from the band to the hopping conduction in the film. The observed transition temperature decreases with the increasing boron content of the film.

However, when 2000 ppm of nitrogen was introduced in feed gas, growth rate was double than that using 1000 ppm of nitrogen but some non-epitaxial growth was found on the surface of doped film. The color of seed crystal changed from yellow to pale blue then to dark blue as the boron concentration in the film increased. A few additional bands along with first order Raman line were visible in the lower wavenumber region in Raman spectrum of boron-doped diamond films. The downshift and broadening of Raman line was also observed with increasing boron content in the crystal. The decrease in width and upward shift of Raman line was also observed in the samples that grew in presence of nitrogen. The growth rate of diamond was observed to be decreasing with increasing boron content in the film. Temperature dependent resistivity measurements showed that the current conduction mechanism depends upon the doping level and obeys semiconductor behavior in the experimental temperature range 140 to 600 K.

### ACKNOWLEDGMENT

This material is based upon work supported by the Department of Energy National Nuclear Security Administration under Award Number DE-NA0002014.